# How relevant is the predictive power of the h-index? A case study of the time-dependent Hirsch index.

**Michael Schreiber**

*Institute of Physics, Chemnitz University of Technology, 09107 Chemnitz, Germany.
Phone: +49 371 531 21910, Fax: +49 371 531 21919
E-mail: schreiber@physik.tu-chemnitz.de*

The h-index has been shown to have predictive power. Here I report results of an empirical study showing that the increase of the h-index with time often depends for a long time on citations to rather old publications. This inert behavior of the h-index means that it is difficult to use it as a measure for predicting future scientific output.

## 1. Introduction

The Hirsch index or h-index of a researcher has been defined by Hirsch (2005) as the largest number $h$ of publications of a researcher which have obtained at least $h$ citations each. It has become popular and is frequently used as a measure of the scientific achievement of a researcher. But it is controversial whether the h-index is indeed a representative measure (Lehmann, Jackson, & Lautrup, 2007; Leeuwen, 2008; Bornmann, 2012). Hirsch (2007) has shown that the h-index has predictive power in the sense, that there is a high correlation between the Hirsch index values after 12 years and after 24 years of the career of researchers. In short, "the h-index is a better predictor of itself than any of the other indicators" investigated in that study and "the h-index has the highest ability to predict *future* scientific achievement". Hirsch concludes that the h-index "can be profitably used in academic appointment processes and to allocate research resources".

It is the purpose of the present investigation to raise doubts about this conclusion. Using empirical data, I demonstrate that the increase of the h-index with time after a given point of time (e.g. the time of appointment or the time of allocating resources) is not necessarily related to the scientific achievements after this date. Specifically, I show examples where the growth of the h-index is the same, irrespective of whether the investigated researcher had performed as he or she did or whether he (she) had not published any further work.

Of course, the above mentioned consideration that the h-index is used in academic hiring processes and research fund allocation processes is only one aspect of looking at the evolution of the h-index with time. The other aspect is that a substantial increase of the h-index even after the researcher has stopped publishing bears testimony of the significance of this researcher's achievements.

Recently, Acuna, Allesina, and Kording (2012) have fitted the time-dependent h-index of neuroscientists using a large number of 18 parameters and were able to predict the future h-index rather accurately for several years. Even a reduced set of 5 parameters and thus 6 coefficients yielded a surprisingly good fit. However, the warning that "with four parameters I can fit an elephant, and with five I can make him wiggle his trunk" which has been attributed to von Neumann by Enrico Fermi (Dyson, 2004) indicates that such a fit can easily suggest an accuracy of the prediction which may not be meaningful. Rousseau and Hu (2012) have already raised objections against this approach of predicting the h-index by a complicated fit.

## 2. The first example

For the following investigations the citation records were determined in the ISI Web of Science database in July and October 2012. Great care has been taken to establish the integrity of the datasets with respect to homonyms, excluding other authors with the same name as the investigated researchers. I note that the ISI author-finder tool turned out not to be helpful at least for my own dataset. The results have been downloaded from the citation report into a spreadsheet where it is straightforward to sum the citations up to a given year and to count the papers with high citation frequencies up to the value of $h$ also selectively, namely depending on the publication year interval as desired. In this way the h-index can easily be determined for year $y$ considering (only) publications up to a certain year $s \leq y$. I shall denote the thus obtained index as $h_s(y)$ in the following. It signifies the h-index for the year $y$, if the researcher had stopped publishing in the year $s$. Obviously, for $s = y$ one obtains the usual h-index: $h = h_y(y)$. Values $s > y$ are not meaningful.

Out of personal interest I had determined the time evolution of the h-index for myself, as displayed in figure 1, which shows a more or less steady increase with a slope of 1 for 30 years which is a rather satisfactory value. In the earlier years after my first publication as a diploma student in 1976 the increase was somewhat slower.

I have then asked myself the question how my h-index would have evolved, if I had stopped publishing, e.g., in 1983 shortly after my post-doc stay. As shown in figure 1, the index $h_{1983}(y)$ would have been the same as $h_y(y)$ for the next 4 years $y$ and would have deviated only two index points after 6 years. Likewise, if I had become lazy after obtaining my habilitation degree in 1987, then $h_{1987}(y) = h_y(y)$ until 1991 and $h_{1987}(1993) = h_{1993}(1993) - 2$. Similarly, if I had stopped publishing in 1990, the year of my first appointment to a tenured professorship, then after 6 years my value of $h_{1990}(y)$ would also have been only two points smaller than $h_{1996}(1996)$. A somewhat longer time of inert behavior of my h-index can be observed starting in 1994. If I had stopped research in 1994 shortly after my promotion to full professorship, 6 years later my index $h_{1994}(2000)$ would have been the same as $h_{2000}(2000)$. Even more drastic is the effect if I had retired already in 1998. For the first 4 years my h-index growth would not have suffered at all, and even after 12 years it would have been reduced only by one index point, i.e. $h_{1998}(2010) = h_{2010}(2010) - 1$. But if I had chosen to stop working in 2002, then there would have been a deviation only in 2009 and 2012 by a single index point. And since 2006 no change can be detected, i.e., $h_s(y) = h_y(y)$ for $s \geq 2006$.

In conclusion, for all the mentioned years $s$ it would have taken several years until one could have detected by means of the h-index that I had completely stopped publishing. Thus by means of the h-index one cannot distinguish for a long time after the decision year $s$ whether I lived up to the expectations in the year $s$ or whether I stopped working. The h-index increased after the year $s$ not because of my performance during that time span, but because of past achievements, namely that my previous publications were still frequently cited.

## 3. Further examples

In order to see whether my own case might be an exception, I have analyzed the citation record of J. Hirsch, see figure 2. He started publishing in 1976, too. His index values are up to a factor of two higher than mine, and the inertness of the index is even stronger. This might be related to the fact that it is more difficult for additional publications to contribute to the h-index, if the index is already rather high. First, figure 2 shows that the index growth would have been the same after three years, if Hirsch had stopped publishing in 1983, but already one year later a deviation of 5 index points can be detected. However, assuming that Hirsch had not published any more papers after 1989, then even 10 years later the index growth would have been unchanged, i.e. $h_{1989}(1999) = h_{1999}(1999)$ and after 16 years $h_{1989}(2005) = h_{2005}(2005) - 2$ only. For $s = 1992$, even after 13 years the same h-index increase would be observed,

$h_{1992}(2005) = h_{2005}(2005)$ and after 16 years $h_{1992}(2008) = h_{2008}(2008)$ - 2. If Hirsch had stopped working in 2001, his index would have been unaffected in 2010 and even in 2012 deviate only by one index point. From 2005 onwards no change would have resulted except a deviation of one index point in the year 2009.

As further examples I present in figures 3 and 4 the corresponding results for M. Cardona and E. Witten. I have chosen these two physicists, because Hirsch (2007) had also singled out these researchers. Incidentally, Witten also started publishing in 1976. In his case, the inert behavior of the h-index is relatively short, even though his h-index values are more than twice as high as Hirsch's values and about 4 times as high as my values. As depicted in figure 3, for $s = 1986$ the index growth would have remained unchanged for only 3 years, in the case $s = 1991$ for only 5 years, and in the case $s = 1999$ for 7 years. Only recently the delay in the response would have become larger: if he had stopped publishing in 2002, the index would have been unimpaired in 2008, and deviate by only one index point in 2012. From $s = 2005$ onwards no change at all can be detected. In comparison with figures 1 and 2 the time spans during which the h-index remains unchanged independent of the publication performance are overall shorter. This observation indicates that the assumption made in the discussion of Hirsch's data, namely that the index is more inert for higher index values cannot be generalized.

Looking at the data for M. Cardona in figure 4, one sees also impressively high values of the h-index, even considering that he started publishing 18 years earlier than the other researchers discussed in the previous paragraphs. If he had stopped publishing in 1980, his h-index would have grown unchanged until 1988. In that year a distinct kink in the h-index curve can be seen. It is interesting to note that the $s = 1980$ curve continues its pre-1988 increase with the same slope until 1996 and with two exceptions even until 2001. This means that an extremely inert behavior of $h_{1980}(y)$ as a measure of achievement based on the performance until 1980 is observed until 2001, i.e. for 20 years. A rather inert behavior can also be seen for $s = 1990$, because after 10 years $h_{1990}(2000) = h_{2000}(2000)$ and even after 16 years $h_{1990}(2006) = h_{2006}(2006) - 2$ only. Similarly, if M. Cardona had stopped publishing in 1994, the h-index would have grown unchanged for 12 years and decreased by one index point only in the subsequent 2 years. Likewise, for $s = 1997$, after 12 years we find $h_{1997}(2009) = h_{2009}(2009)$. Stopping publications in the year of his formal retirement, 2000, would also have had no impact on his h-index curve for 11 years until 2011 and only a slight reduction of one index point in 2012. For $s \geq 2005$ no change at all can be detected, very similar to the other examples.

### 4. Conclusions

Hirsch (2007) has demonstrated the predictive power of the h-index by showing that "a researcher with a high h-index after 12 years is highly likely to have a high h-index after 24 years". In the present investigation I have chosen different time intervals. In the presented examples the h-index increases rather smoothly with time (with one exception: Cardona 1988). This confirms the claim that the h-index is a good predictor of itself. It is therefore tempting to choose the h-index in appointment processes or for the purpose of the allocating resources. This temptation is based on the assumption that the predicted increase of the h-index is correlated with the expected future performance of a candidate. However, this may not be true.

As demonstrated in 4 examples the increase of the h-index does not necessarily depend on the factual performance for several years in the future, but is more likely to result from previous, often rather old publications, i.e., on passée achievements. Thus even if one does not perform as expected, one would not disappoint those evaluators who measure the performance in terms of the h-index after 3 or 5 years or even later. I admit that for the discussion above I have selected the most impressive examples for the year $s$ so that the times of inertia are particularly long and not representative. Nevertheless, these examples demonstrate that it is dangerous to draw conclusions from the predictive value of the h-index with respect to future performance.

It would be cynical to consider it an advantage of the h-index that one might stop working without administrators noting it when they are looking only at the h-index. But I can even imagine bureaucrats, who do not even care, as long as they can have their expectations fulfilled when measuring the expected increase of the h-index irrespective of whether the expected performance has really been achieved or not. Although this is certainly not a new finding, I conclude that one should be very careful when using indicators like the h-index for the prediction of future scientific performance.

Only 4 examples have been presented in the present investigation, but in my opinion this is sufficient to raise doubts concerning the idea that the h-index can be used to predict future scientific achievements. I have not yet investigated other examples, but I expect similar results in many, probably most cases. Certainly more comprehensive future studies are necessary to validate the results, not only using larger datasets, but also comparing scientists from different disciplines and at different scientific age. For this purpose, one should also define a quantitative measure in order to calculate the inertia of the h-index for a given time span. For example, one might utilize the largest duration $d(y) = y - s$ for which the h-index remains unchanged or only slightly changed, like $\Delta = h_y(y) - h_s(y) = 0$, or 1, or 2. This would be a function of $y$ and would need averaging (possibly excluding an initial time span avoiding problems with very small index numbers and problems with transient behavior of the index at the beginning of a career) or some other kind of condensing the information. Another possibility would be to fix the duration time $d = y - s$ at, say, 5 or 10 years and then determine the deviation $\Delta(y)$ and analyze this function, again after some kind of averaging or otherwise condensing the information. In conclusion, only with a variety of future studies the predictive power of the h-index might be assessed more clearly.

**Figure 1.** Time evolution of the h-index for the publications of the present author (top line). Additionally the dependence of $h_s(y)$ is shown for selected years (see legend) starting with the year $s$ which is assumed as the year in which the author had stopped publishing.

**Figure 2.** Same as figure 1, but for J. Hirsch.
**Figure 3.** Same as figure 1, but for E. Witten.
**Figure 4.** Same as figure 1, but for M. Cardona.

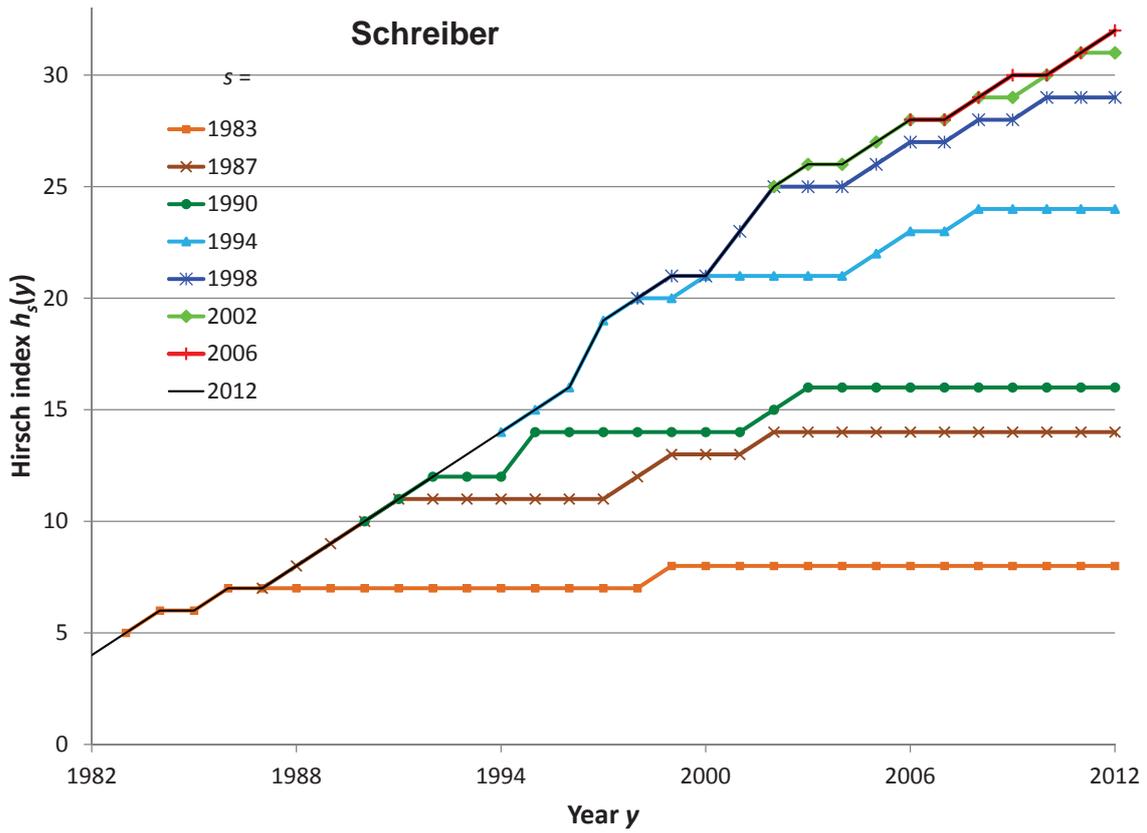

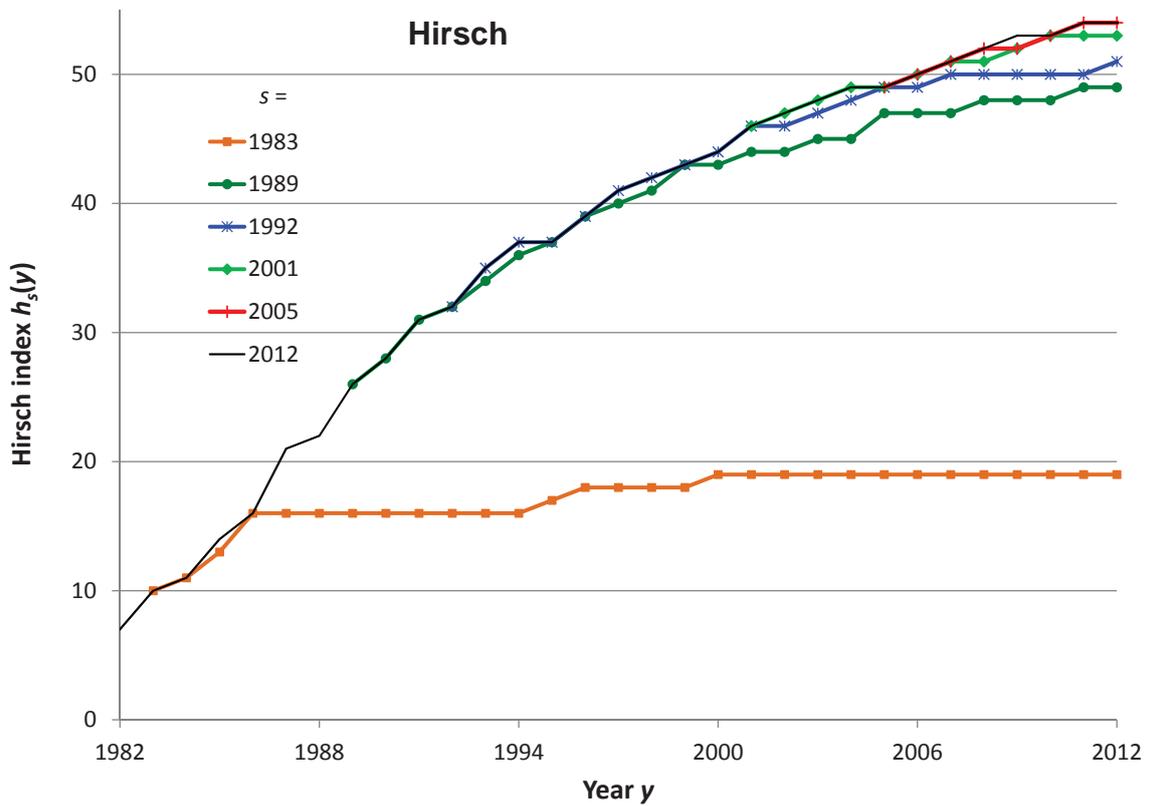

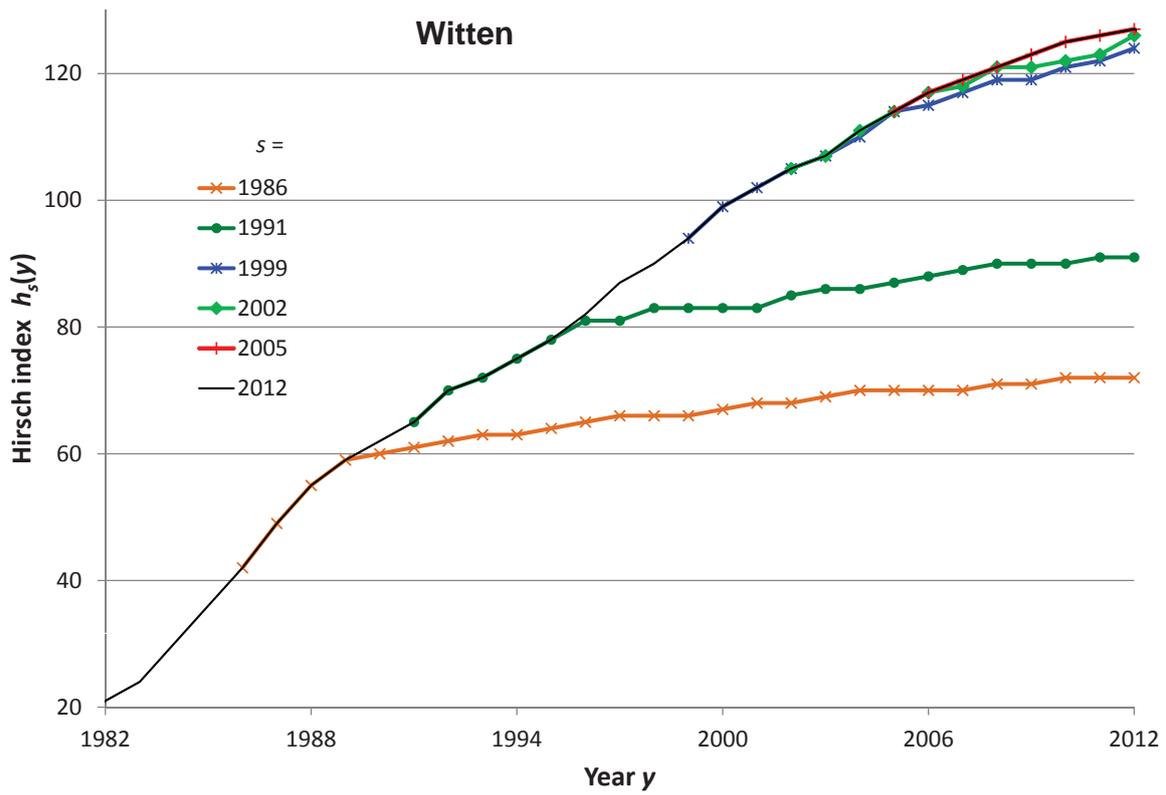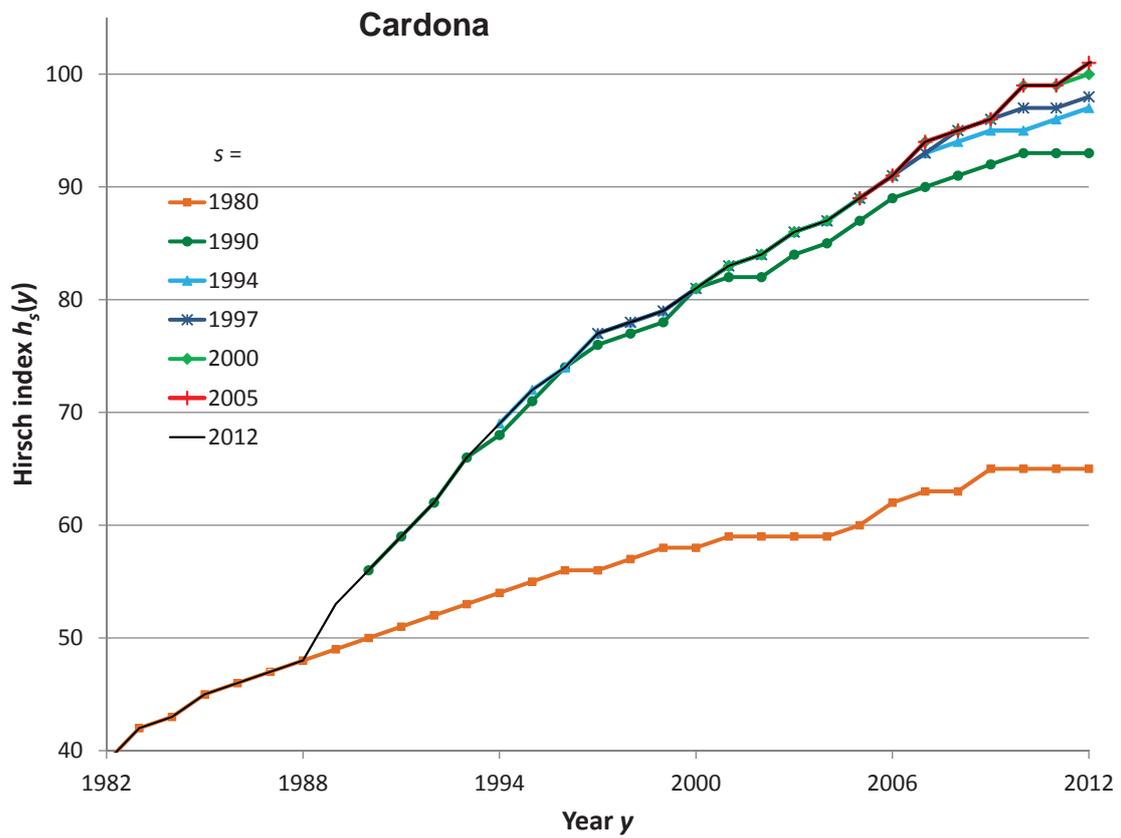